(technical note)

Multiple Protein Profiler 1.0 (MPP): A webserver for predicting and visualizing physiochemical properties of proteins at the proteome level.

Authors
Gustavo Sganzerla Martinez[1,2,3], Mansi Dutt[1,2,3], Anuj Kumar[1,2,3], David J Kelvin[1,2,3*]

Affiliations
[1]Department of Microbiology and Immunology, Dalhousie University, Halifax, Nova Scotia, Canada, B3H4H7.
[2]Department of Pediatrics, Izaak Walton Killam (IWK) Health Center, Canadian Center for Vaccinology (CCfV), Halifax, Nova Scotia, Canada. B3H4H7.
[3]BioForge Canada Limited. Halifax, NS, Canada.

*Correspondence: David J Kelvin (David.Kelvin@dal.ca)

Abstract

Determining the physicochemical properties of a protein can reveal important insights in their structure, biological functions, stability, and interactions with other molecules. Although tools for computing properties of proteins already existed, we could not find a comprehensive tool that enables the calculations of multiple properties for multiple input proteins on the proteome level at once. Facing this limitation, we have developed Multiple Protein Profiler (MPP) 1.0 as an integrated tool that allows the profiling of 12 individual properties of multiple proteins in a significant manner. MPP provides a tabular and graphic visualization of properties of multiple proteins. The tool is freely accessible at https://mproteinprofiler.microbiologyandimmunology.dal.ca/

**Introduction**

Proteins are the ubiquitous biological macromolecules that play principal roles in living organisms. These large and complex molecules have been well studied for their involvement in the structure, function, and regulation of tissues and organs [1]. In addition, these molecules are important structural components that carry out a plethora of biological functions, from replicating DNA and transporting/storing molecules to catalyzing metabolic pathways and much more [2]. Twenty different amino acids significantly contribute to the formation of proteins in the body. Antibodies, enzymes, hormones, messengers, structural components, and transport/storage molecules are among the popular examples of proteins. At the structural level, proteins can be defined by their primary, secondary, tertiary, and quaternary structures [3,4]. Physiochemical properties of the proteins play a vital role in determining their structures, biological functions, stability, and molecular interactions with other micro- and macromolecules [5]. At the sequence level, a set of properties including molecular weight, theoretical isoelectric point (pI), estimated half-life, instability index, amino acid composition, atomic composition, aliphatic index, and grand average of hydropathicity (GRAVY) index etc., are defined as the popular physiochemical properties of the proteins. These properties are crucial to understand and annotate the complex biological process [6-8]. So far, a number of tools and databases are available in the public domain for predicting the physiochemical properties of proteins, including ExPASy ProtParam (https://web.expasy.org/protparam/), AAindex (https://www.genome.jp/aaindex/); PeptideMass (https://web.expasy.org/peptide_mass/); and Biopython (https://biopython.org/) etc. Recently, [9] developed the GUD-VE, a graphical user interface (GUI) tool for visualizing the physicochemical properties of proteins. However, no web server is available for the prediction of physiochemical properties of a bulk of proteins or whole proteome. We therefore developed a user-friendly webtool, the Multiple Protein Profiler 1.0 (MPP) (http://mproteinprofiler.microbiologyandimmunology.dal.ca/), to simultaneously predict the physiochemical properties of either a bulk of proteins or whole proteomes. This webserver also facilitates end users in rendering predicted results in the form of attractive ready-to-publish graphs. The MPP web server is, to our knowledge, the first of its kind. We anticipate that the MPP will be useful for the scientific community to predict and annotate the biological functions of the proteins.

**Development**

The Multiple Protein Profiler (MPP) tool was developed using Python version 3.10.5 in the backend and Hypertext Markup Language (HTML) 5 in the frontend; MPP is implemented as a web application using the Django framework. Since MPP is not platform/browser dependent, it is able to run in both desktop and mobile versions of Google Chrome, Mozilla Firefox, Apple's Safari, Opera, and Microsoft Edge. No database is required for MPP to run; therefore, no information on the input data of the users is stored. Also, MPP is designed so that no login is required (**Figure 1**).

To locally run some of the functionalities employed by MPP, the following dependencies are required: the Python modules SeqIO and ProteinAnalysis from the packages Bio and Bio.SeqUtils.ProtParam, respectively; the modules StringIO and BytesIO from the package io; and the re, csv, and matplotlib packages.

The application of MPP is hosted at a server located in Central Canada (Toronto) with the IP address 172.105.99.100. The Django-ready server runs the Debian 10 Linux distribution.

The deployment of MPP to its remote server, and consequently to the web for public use, is linked to a cloned Git repository, which is public at https://github.com/gustavsganzerla/mpp. A local version of GitHub (version 2.20.1) is installed at the remote server. Any subsequent changes to the

MPP application are firstly locally implemented in a production server, then committed to the main Git repository, and finally pulled from the repository by the remote server.

**Supported operations**

MPP is able to read a collection of proteins as an uploaded .fasta file or manually input proteins in a textbox and return a list of computed properties of each protein. On its 1.0 iteration, MPP can perform the following tasks for each protein: *i*) determine the length; *ii*) calculate the GRAVY; *iii*) calculate the aliphatic index; *iv*) calculate the instability index; *v*) determine whether the protein is stable or unstable; *vi*) calculate the molecular weight; *vii*) calculate the aromaticity; *viii*) calculate the isoelectric point; *ix*) determine the charge at pH 7; *xi*) determine the fraction of secondary structure; *xii*) determine the molar extinction coefficient; xiii) calculate the amino acid composition and; *xiv*) calculate the atomic composition. Moreover, for the properties *i* to *ix*, MPP can provide a visual plot encompassing all the input proteins. Below, we show the description of each property calculated by MPP.

- i) Length: the total amino acids composing each input protein are calculated. An integer value is returned. Each protein gets converted to a string and the *len()* function is applied. This property counts with a plotting function, in which a histogram is provided for checking the frequency of the length of all input proteins.
- ii) GRAVY: each amino acid composing an input protein has a hydropathicity value, previously calculated by [10] Kyte and Doolittle, 1982 (the reference list is available at Supplementary Table S1); the GRAVY function of MPP sums the hydropathicity values of a protein and divides it by the length of the protein, thus returning a score. This property counts with a plotting function, in which a scatter plot is provided for each individual GRAVY score calculated.
- iii) Aliphatic index: calculates the aliphatic index of a protein sequence by first counting the occurrences of specific aliphatic (non-polar) amino acids (Alanine, Valine, Isoleucine, and Leucine), then computing their mole percentages within the protein. Using predefined constants 'a' and 'b,' it calculates the aliphatic index, which reflects the hydrophobicity of the protein. This index is based on the composition of these aliphatic amino acids and their contribution to the overall structure, with higher values indicating a higher aliphatic character in the protein sequence. The calculation was based on [11]. This property counts with a plotting function, in which a scatter plot is provided for each individual aliphatic index score calculated.
- iv) Instability index: for calculating the instability index of input proteins, we used the Bio.SeqUtils.ProtParam module from BioPython. The function *instability_index(self)* calculates the stability of proteins. An input protein needs first to be read as a ProteinAnalysis class to be used as input parameter for the *instability_index(self)* function. This property counts with a plotting function, in which a scatter plot is provided for each individual instability index score calculated. In addition, each plot generated by MPP's calculation of the instability index will have a horizontal line in the instability value (y-axis) of 40. Proteins with the instability index above 40 indicate the protein is unstable and has a short half-life [12].

v) Stability: the threshold of 40 is accounted for determining the stability of each protein. Values equal to or lesser than 40 are defined as stable while values above 40 are defined as stable. This property counts with a plotting function, in which a pie plot is generated accounting the proportion of stable and unstable proteins for a given input dataset.

vi) Molecular weight: for calculating the molecular weight of input proteins, we used the Bio.SeqUtils.ProtParam module from BioPython. The function *molecular_weight(self)* takes a protein read as a ProteinAnalysis object as input and returns the sum of the atomic weight of all atoms in the chemical structure of the protein. This value is represented in Daltons (Da). This property counts with a plotting function, in which a scatter plot is provided for each individual molecular weight calculated.

vii) Aromaticity: for calculating the aromaticity of input proteins, we used the Bio.SeqUtils.ProtParam module from BioPython. The function *aromaticity(self)* takes a protein read as a ProteinAnalysis object as input and returns the relative frequency of aromatic amino acids (i.e., phe, trp, and tyr) that compose the input proteins. This property counts with a plotting function, in which a scatter plot is provided for each individual aromaticity.

viii) Isoelectric point: for calculating the isoelectric point, we used the Bio.SeqUtils.ProtParam module from BioPython. The function *isoelectric_point(self)* takes a protein read as a ProteinAnalysis object as input and returns the pI of a protein, i.e., the pH in which the net charge of a protein molecule is zero. This property counts with a plotting function, in which a scatter plot is provided for each individual pI.

ix) Secondary structure fraction: for calculating the fraction of the helix, turn, and sheet of a protein, we used the Bio.SeqUtils.ProtParam module from BioPython. The *secondary_structure_fraction(self)* function takes as input a protein read as a ProteinAnalysis object and returns the fraction of amino acids that tend to be in the helix, turn, and sheet of the protein. The amino acids V, I, Y, F, W, and L are in the helix. The amino acids N, P, G, and S are in the turn. The amino acids E, M, A, and L are in the sheet.

x) Molar extinction coefficient: for calculating the molar extinction coefficient of each input protein, we used the Bio.SeqUtils.ProtParam module from BioPython. The method *molar_extinction_coefficient(self)* returns the molar extinction coefficient of proteins containing cysteines and cystine (a dimer of cysteine residues connected by a disulfide bond).

xi) Atomic composition: for determining the atomic composition of the input protein, we accounted for the number of carbon (C), hydrogen (H), oxygen (O), nitrogen (N), and sulfur (S) atoms present in each amino acid composing each input protein. A chemical equation of the respective number of CHONS was returned. The number of CHONS molecules per each amino acid were taken from [3] and are available in Supplementary Table S2.

Apart from the beforementioned calculation of properties of proteins, MPP also generates a downloadable .CSV file containing the accession number, protein sequence, protein description, length, GRAVY, aliphatic index, instability index, stability, molecular weight, aromaticity, isoelectric point, and charge at pH 7 of each input protein (**Figure 2**).

Finally, MPP also enables the individual visualization of each input protein. In the results page, we included an URL in the accession number (first column) of each input protein that redirects the user to a subsequent page, displaying the accession number, description, amino acid sequence, length, GRAVY, aliphatic index, instability index, stability, molecular weight, charge at pH 7, atomic composition, and percentage of each amino acid of the selected protein.

**Comparison with other tools**

A common webtool for the computation of physical properties of proteins is ProtParam from the Expasy server [13]. In fact, some of the functionalities we applied at MPP are provided in the ProtParam Python module. However, ProtParam is limited to profiling one protein at a time; it is not able to profile entire proteomes and/or lists of proteins in a single run. We acknowledge that the developers of ProtParam, by enabling their Python module, could adapt the code to iteratively profile multiple proteins at once. However, Python programming proficiency on the user side would be required to do so.

Moreover, we found the webserver bioinformatics.org [14], which allows the individual calculation of GRAVY, molecular weight, and isoelectric point of multiple proteins at once. This tool is not able to streamline multiple properties, requiring one run to be executed for each property.

**Concluding remarks**

In developing MPP, we aimed to provide a webtool that allows users to input entire proteomes or custom-made lists of proteins for profiling by our tool. The development philosophy we applied for building MPP allows us to iteratively add new features to our tool. Finally, the transparent development of MPP, in which the source code is publicly available on a repository, welcomes interested users and developers to contribute to the lifecycle of MPP.

**Conflict of Interest**

The authors GSM, AK, MD, and DJK are members of the company BioForge Canada Limited. BioForge Canada Limited is a company that uses bioinformatics in immunological approaches in the monitoring, prevention, and treatment of infectious diseases. The authors disclose that the interests of BioForge Canada Limited had no impact in this study.

**Acknowledgements**

The corresponding author (DJK) is the Canada Research Chair in Translational Vaccinology and Inflammation. The authors are also thankful to Dr. Nikki Kelvin for her key inputs in proofreading and getting this work ready for publication.

**Funding**

This work was supported by awards from the Canadian Institutes of Health Research, the Mpox Rapid Research Funding initiative (CIHR MZ1 187236), Research Nova Scotia (DJK), Atlantic Genome/Genome Canada (DJK), Li-Ka Shing Foundation (DJK), Dalhousie Medical Research Foundation (DJK).

Figures

**Figure 1.** Homepage of the Multiple protein Profiler with the options of sample input and upload files in fasta format.

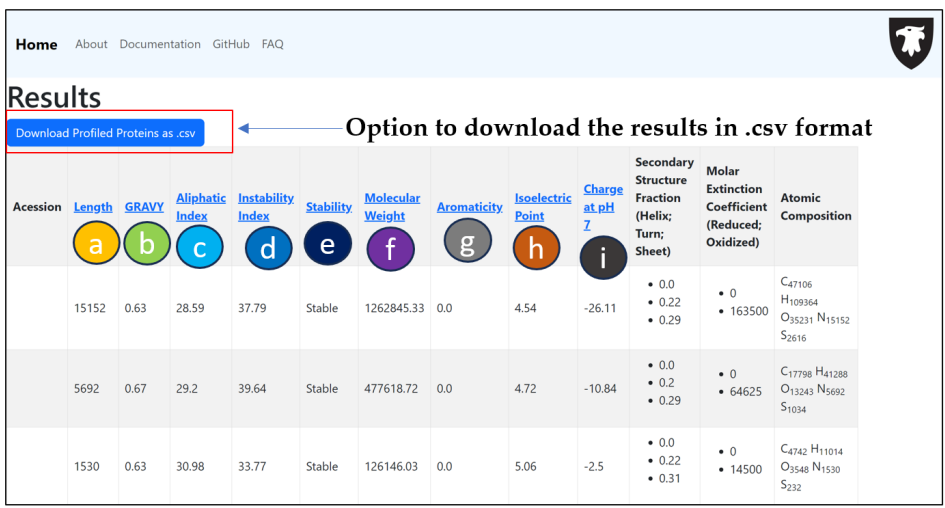
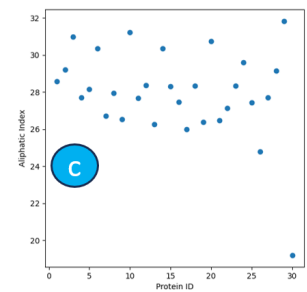
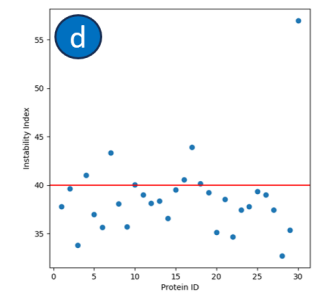
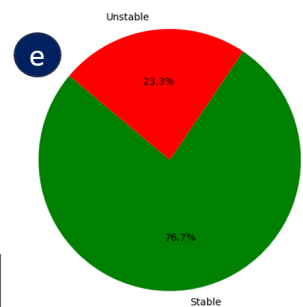
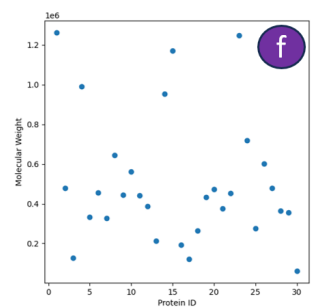
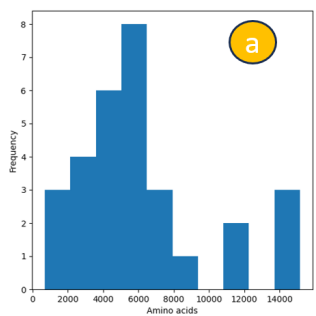
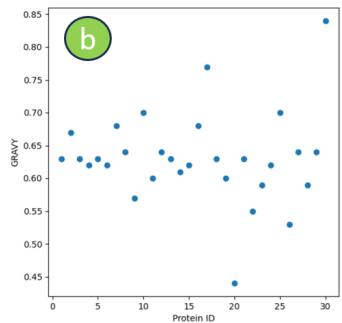
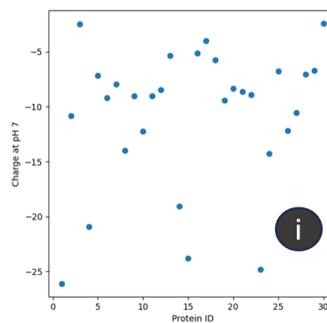
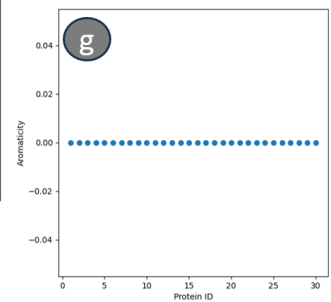
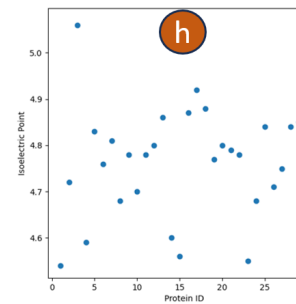

**Figure 2. Graphical representation of MPP output for bulk sequences or whole proteome**: (a) Lenth of amino acids; (b) GRAVY; (c) Aliphatic index; (d) Instability index; (e) Stability; (f) Molecular weight; (g) Aromaticity; (h) Isoelectric point; and (i) Charge at pH 7.